\documentclass[sigconf, screen]{acmart}

\usepackage{algorithm}
\usepackage{algorithmic}
\usepackage{multirow}
\usepackage{graphicx}
\usepackage{subfigure}
\usepackage{bbold} 
\usepackage{mathtools}
\usepackage{enumitem}
\usepackage{balance}

\AtBeginDocument{%
  \providecommand\BibTeX{{%
    \normalfont B\kern-0.5em{\scshape i\kern-0.25em b}\kern-0.8em\TeX}}}

\copyrightyear{2021}
\acmYear{2021}
\setcopyright{acmcopyright}\acmConference[MM '21]{Proceedings of the 29th ACM International Conference on Multimedia}{October 20--24, 2021}{Virtual Event, China} \acmBooktitle{Proceedings of the 29th ACM International Conference on Multimedia (MM '21), October 20--24, 2021, Virtual Event, China}
\acmPrice{15.00}
\acmDOI{10.1145/3474085.3475195}
\acmISBN{978-1-4503-8651-7/21/10}

\begin{document}

\fancyhead{} 

\title{Video Background Music Generation \\ with Controllable Music Transformer}

\author{Shangzhe Di}
\authornote{Both authors contributed equally to this research.}
\affiliation{%
  \institution{Beihang University}
  \city{Beijing}
  \country{China}
}
\email{dishangzhe@buaa.edu.cn}

\author{Zeren Jiang}
\authornotemark[1]
\affiliation{%
    \institution{Beihang University}
    \city{Beijing}
    \country{China}
}
\email{zeren.jiang99@gmail.com}

\author{Si Liu}
\authornote{Corresponding author.}
\affiliation{%
    \institution{Beihang University}
    \city{Beijing}
    \country{China}
}
\email{liusi@buaa.edu.cn}

\author{Zhaokai Wang}
\affiliation{%
    \institution{Beihang University}
    \city{Beijing}
    \country{China}
}
\email{wzk1015@buaa.edu.cn}

\author{Leyan Zhu}
\affiliation{%
    \institution{Beihang University}
    \city{Beijing}
    \country{China}
}
\email{leyan.zhu@buaa.edu.cn}

\author{Zexin He}
\affiliation{%
    \institution{Beihang University}
    \city{Beijing}
    \country{China}
}
\email{jacquesdeh@buaa.edu.cn}

\author{Hongming Liu}
\affiliation{%
    \institution{Charterhouse School}
    \country{Godalming, Surrey, UK}  
}
\email{lclcpg2018@gmail.com}

\author{Shuicheng Yan}
\affiliation{%
    \institution{Sea AI Lab}
    \country{Singapore}
}
\email{Yansc@sea.com}

\renewcommand{\shortauthors}{Di and Jiang, et al.}

\begin{abstract}
In this work, we address the task of video background music generation. Some previous works achieve effective music generation but are unable to generate melodious music tailored to a particular video, and none of them considers the video-music rhythmic consistency. To generate the background music that matches the given video, we first establish the rhythmic relations between video and background music. In particular, we connect timing, motion speed, and motion saliency from video with beat, simu-note density, and simu-note strength from music, respectively. We then propose CMT, a \textbf{C}ontrollable \textbf{M}usic \textbf{T}ransformer that enables local control of the aforementioned rhythmic features and global control of the music genre and instruments. Objective and subjective evaluations show that the generated background music has achieved satisfactory compatibility with the input videos, and at the same time, impressive music quality. Code and models are available at \url{https://github.com/wzk1015/video-bgm-generation}.

\end{abstract}




\maketitle

\begin{figure}[htbp]
  \centering
  \includegraphics[width = 0.96\linewidth]{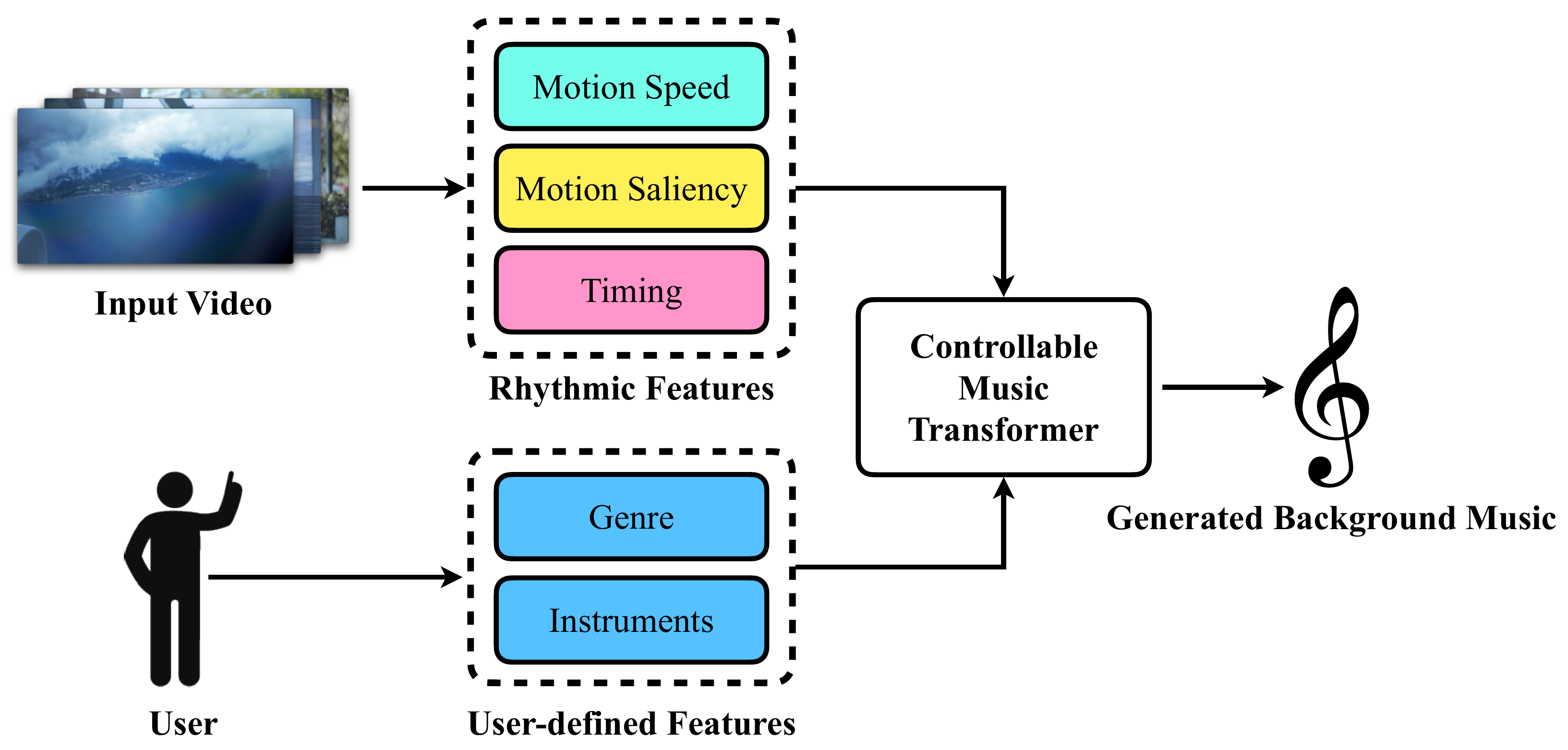} 
    \caption{An overview of our proposed framework. We establish three rhythmic relations between video and music. Together with the user-defined genre and instruments, the extracted rhythmic features are then passed to a carefully designed Controllable Music Transformer to generate proper background music for a given video.}
  \label{overview}
\end{figure}
\vspace{-0.3cm}

\section{Introduction}

Nowadays, people can conveniently edit short videos with video editing tools and share their productions with others on social video-sharing platforms. 
To make a video more attractive, adding background music is a common practice, which, however, is not so easy for those without much knowledge of music or film editing.
In many cases, finding proper music material and make adjustments to make the music fit for a video is already very difficult and time-consuming. Not to mention the copyright protection issue that is causing increasingly broader public concern.
Therefore, automatically generating appropriate background music for a given video becomes a task of real-world significance, yet it is barely studied in the multimedia community to the best of our knowledge.
There are some previous works~\cite{midinet} \cite{MuseGan} \cite{music_transformer} tackling music generation based on deep learning models,
However, none of them takes into account the music-video rhythmic consistency. 

In this paper, we address the task of video background music generation. 
Instead of adopting a costly data-driven practice as in tradition, \emph{i.e.} collecting paired video and music samples for training models, we explore the rhythmic relations between video and background music and propose a transformer-based method free of reliance upon annotated training data. In Fig.~\ref{overview}, we provide an illustration of this task and an overview of our method.

A video may contain diverse visual motions of different patterns.
For example, a man in a video is walking, fast or slowly, and he may suddenly stop. 
To generate proper background music for such a scene, we should consider the speed and change of the motion.
In particular, we establish three rhythmic relations between the video and background music.
Firstly, for a video clip, fast motion (e.g., in a sport ) should correspond to intense music, and vice versa. 
Accordingly, we build a positive correlation between \textbf{motion speed and {simu-note density}}, where motion speed refers to the magnitude of motion in a small video clip calculated by the average optical flow, and simu-note density is the number of simu-notes per bar. 
A \emph{simu-note} \cite{pianotree} is a group of notes that start simultaneously, as shown in Fig.~\ref{illustration}. 
Secondly, distinctive motion changes,  such as shot boundaries, should correspond to strong musical beats or music boundaries, making the audience feel both visual and auditory impact at the same time, leading to a more rhythmic video.
Therefore, we align the \textbf{local-maximum motion saliency with simu-note strength}, where local-maximum motion saliency labels some rhythmic keyframes and \emph{simu-note strength} is the number of notes in a simu-note. 
Thirdly, it is more harmonious to sync the epilogue and the prologue between the video and the generated background music. 
That is to say; the background music should appear and disappear smoothly along with the start and end of the video. 
Thus, we extract timing information
from the given video, and take it as the position encoding to guide the music generation process, namely \textbf{beat-timing encoding}.

We build our music representation based on compound words~\cite{cp} (CP). 
We group neighboring tokens according to their types to construct 2-D tokens for \emph{note-related} and \emph{rhythm-related} tokens.
These rhythmic features are added as additional attributes in the token. 
Furthermore, we add music genres and used instruments as initial tokens, as shown in the bottom part of Fig.~\ref{overview}, in order to customize the music generation process to match the given video. 
We use linear transformer \cite{linear_transformer} as the backbone of our proposed pipeline to model the music generation process, considering its lightweight and linear-complexity in attention calculation.
During training, we use the Lakh Pianoroll Dataset (LPD)~\cite{MuseGan} to train our model on music modeling, where we provide the above musical features directly.
For inference, the visual features are obtained from the video and used to guide the generation process.

In summary, our contributions are threefold. 
1) For our task \emph{video background music generation}, we propose the Controllable Music Transformer (CMT) model, which makes use of several key relations between video and music, but does not require paired video and music annotated data during training. 2) We introduce new representations of music, including note density and strength of simu-notes, which result in better-generated music and a more controllable multi-track generation process. 
3) Our approach successfully matches the music to the rhythm and mood of a video, and at the same time, achieves high musical quality.
We put a demo video of an input video and our generated music in the supplementary material for demonstrative purposes.

\section{Related Work}

\textbf{Representations of music.} 
Most previous works on symbolic music generation take the music represented in MIDI-like event sequences \cite{music_transformer} \cite{musenet} as input.
REMI \cite{remi} imposes a metrical structure in the input data, \emph{i.e.,} providing explicit notations of bars, beats, chords, and tempo.
This new representation helps to maintain the flexibility of local tempo changes and provides a basis upon which we can control the rhythmic and harmonic structure of the music. 
Compound words \cite{cp} (CP) further converts REMI tokens to a sequence of compound words by grouping neighboring tokens, which greatly reduces the length of the token sequence. 
In this paper, we employ a representation based on CP. 
We categorize music tokens into rhythm-related tokens and note-related tokens. 
We add \emph{genre} and \emph{instrument type} as initial tokens to provide global information of the music, and \emph{density} and \emph{strength} attributes to enable local control of the generation process.

\textbf{Music generation models.} 
Some recent models~\cite{pianotree} \cite{musae} \cite{musicvae} use autoencoders to learn a latent space for symbolic music and generate new pieces. 
Some  \cite{midinet} \cite{MuseGan} consider piano rolls as 2-D images and build models based on convolution networks. 
Since music and language are both represented as sequences, the transformer and its variants are also frequently used as the backbone of music generation models \cite{music_transformer} \cite{remi} \cite{lakhnes} \cite{cp}. 
Apart from generating symbolic music, some models generate audio directly in waveform \cite{wavenet} \cite{gansynth} \cite{samplernn} or indirectly through transcription and synthesis \cite{mastero}. 
Our model is based on linear transformer \cite{linear_transformer}, which implements a linear attention mechanism to reduce time complexity.

\textbf{Composing music from silent videos.}
Previous works on music composition from silent videos focus on generating the music from video clips containing people playing various musical instruments, such as the violin, piano, and guitar \cite{foley_music} \cite{audeo} \cite{multi_instrument_net}. 
Much of the generation result, e.g., the instrument type and even the rhythm, can be directly inferred from the movement of human hands, so the music is to some extent determined.
Comparably, our method works for general videos and aims to produce non-determined generation results.
In addition, there is currently no dataset of paired arbitrary videos and music specifically for this video background music generation task.
In some existing audiovisual datasets like \cite{youtube-8m} \cite{kinetics}, the videos often contain noise like human speech or simply do not involve music.
Due to the lack of annotated training data, traditional supervised training methods based on audiovisual data do not function regarding this task. 
To the best of our knowledge, no existing work focuses on generating music from arbitrary video clips. 
This paper proposes to generate background music based on motion saliency, visual beats, and global timing of the video, along with user-defined music genres and instruments.


\begin{figure*}[t]
    \centering
    \includegraphics[width=0.95\textwidth]{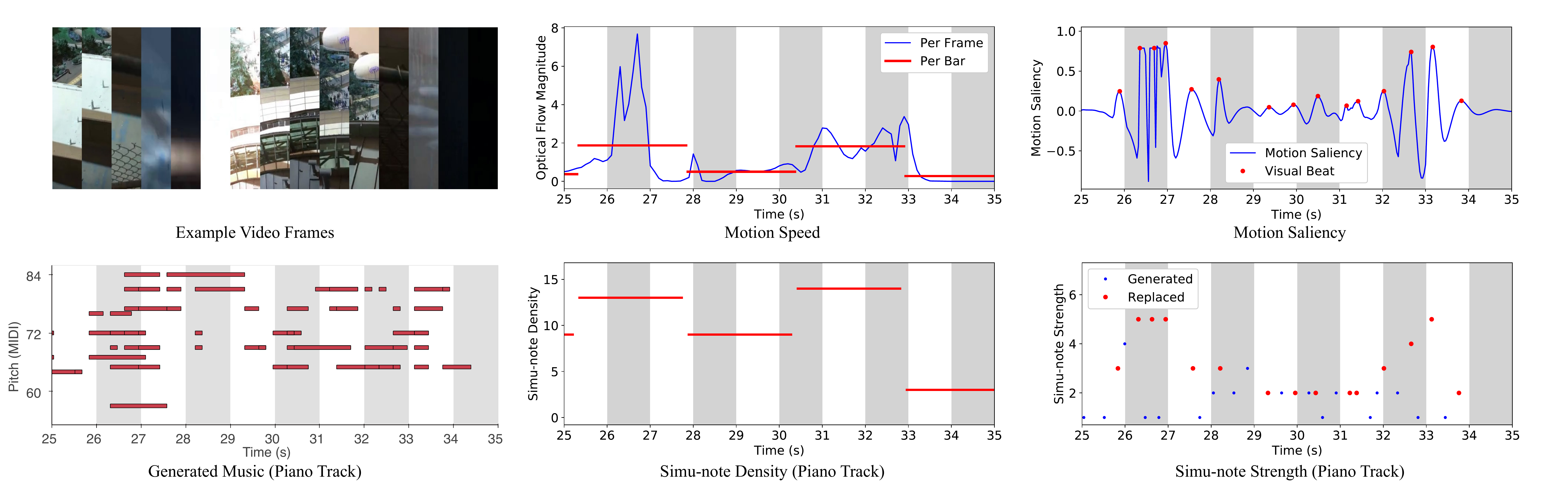}
    \caption{Rhythmic relations between a video clip and the background music generated using our method. Shown here are the original video and its rhythmic features (top), as well as our generated music and its corresponding features (bottom). Our method constructs rhythmic relations between the video and music, and use them to guide the generation of music.}
    \label{fig:firstpage}
\end{figure*}

\section{Establishing Video-Music Rhythmic relations}

One would expect to hear romantic music when watching a romantic film or intense music for a shooting game video. 
Rhythm exists not only in music but also in videos. 
It can reflect how visual motions in a video or note onsets in music are distributed temporally.
To make the generated background music match the given video, we analyze and establish the rhythmic relations between video and music.

Below in Sec.~\ref{sec:time}, we first build a connection between time in the video and musical beat.
Based on this connection, in Sec.~\ref{sec:density}, we establish the relation between motion speed and note density.
In Sec.~\ref{sec:strength}, we build the relation between motion saliency and note strength.
These established relations between video and music will be used to guide the generation of background music that matches rhythmically with the given video.

In the left part of Fig.~\ref{fig:firstpage}, we show a video clip and the generated background music. The generated music has a large simu-note density when motion speed is high (as shown in the middle of Fig.~\ref{fig:firstpage}), and a large simu-note strength when a salient visual beat occurs (as shown in the right part of Fig.~\ref{fig:firstpage}).

\begin{figure}[t]
  \centering
  \includegraphics[width = 0.8\linewidth]{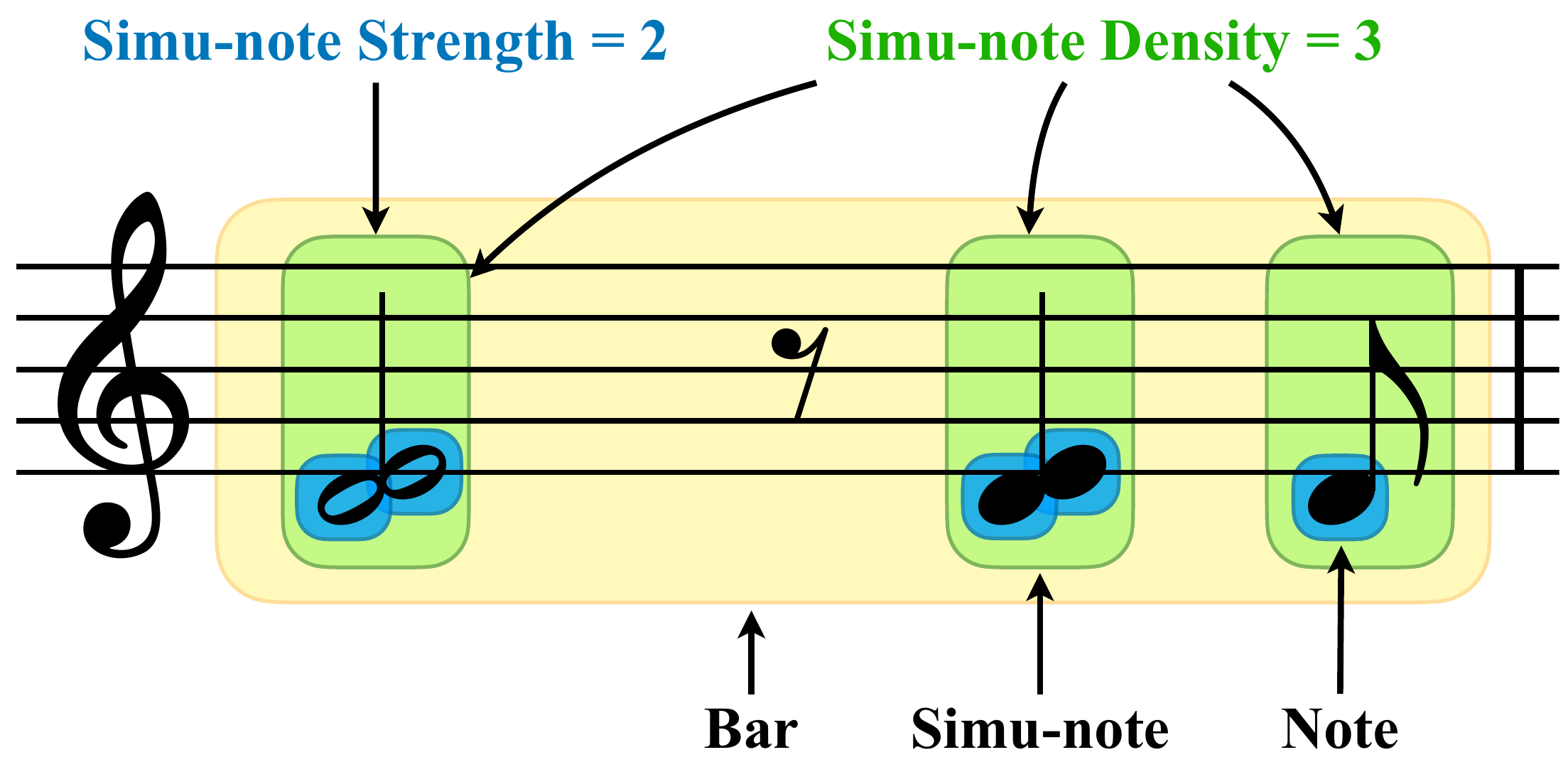} 
  \caption{Illustration of simu-note density and strength. Simu-note density stands for the number of simu-notes in a bar, and simu-note strength stands for the number of notes in a simu-note.}
  \label{illustration}
\end{figure}

\subsection{Video Timing and Music Beat}\label{sec:time}
Formally, we consider a video $V \in \mathbb{R}^{H\times W\times T\times 3}$ that contains $T$ frames. 
We aim to convert the $t$th ($0<t\leqslant T$) frame to its beat number with the following equation:
\begin{equation}
    f_{beat}(t) = \frac{Tempo\cdot t}{FPS\cdot 60},
\end{equation}
where $Tempo$ is the speed at which the background music should be played, and $FPS$ is short for frame per second, which is an intrinsic attribute of the video.
We take $\frac{1}{4}$ beat (one tick) as the shortest unit.

Also, we can convert the $i$th beat to the video frame number based on its inverse function:
\begin{equation}
    f_{frame}(i) = f^{-1}_{beat}(i) = \frac{i\cdot FPS\cdot 60}{Tempo}.
\end{equation}
These two equations serve as the basic block to build the rhythmic relation between the video and music.

\subsection{Motion Speed and Simu-note Density} \label{sec:density}

We first divide the entire video into $M$ clips, with $M$ defined as follows:
\begin{equation}
    M = \left \lceil \frac{f_{beat}(T)}{S} \right \rceil,
\end{equation}
where $T$ is the total number of frames in the video, and we set $S=4$, which means the length of each clip corresponds to 4 beats (one bar) in the music.
Then we calculate the motion speed based on the optical flow in each clip.

Optical flow is a useful tool for analyzing video motions. 
Formally, an optical flow field $f_t(x,y) \in \mathbb{R}^{H\times W\times 2}$ measures the displacement of individual pixels between two consecutive frames $I_t, I_{t+1} \in \mathbb{R}^{H\times W\times 3}$.

In analogy to distance and speed, we define \textbf{\textit{optical flow magnitude}} $F_t$ as the average of the absolute optical flow to measure the motion magnitude in the $t$th frame:
\begin{equation}
    F_{t} = \frac{\sum_{x,y} |f_t(x,y)|}{HW},
\end{equation}
and \textbf{\textit{motion speed}} of the $m$th ($0<m\leqslant M$) video clip as the average optical flow magnitude:
\begin{equation}
    speed_{m} = \frac{\sum_{t=f_{frame}(S(m-1))+1}^{f_{frame}(Sm)} F_t}{f_{frame}(S)}.
\end{equation}

As for music, we manipulate simu-note density to connect with the motion speed. 
Here, a \textbf{\textit{simu-note}}~\cite{pianotree} is a group of notes having the same onset:
\begin{equation}
    \emph{simu-note}_{i,j,k} = \{n_1, n_2, ..., n_N\},
    \label{equ:simu_note}
\end{equation}
where $i$ denotes the $i$th bar, $j$ denotes the $j$th tick (4 ticks is 1 beat) in this bar, $k$ denotes the instrument, and $n$ denotes a single note. 
Compared with notes, the concept of simu-note focuses more on the rhythmic feature since no matter it is a seventh chord or a ninth chord, it is one simu-note.

Moreover, a \textbf{\textit{bar}} can be expressed as a group of non-empty simu-notes:
\begin{equation}
    bar_{i,k} = \{\emph{simu-note}_{i,j,k} | \emph{simu-note}_{i,j,k} \ne \varnothing, j = 1, 2, ..., 16\},
\end{equation}
where $j=1,2,...,16$ as we divide a bar into 16 ticks.
The \textbf{\textit{simu-note density}} of a bar is then defined as:
\begin{equation}
    density_{i} = \left | \left \{ j | \exists k\in \mathbb{K}, \emph{simu-note}_{i,j,k}\in bar_{i,k}  \right \} \right |.
\end{equation}

Then, we statistically analyze the distribution of both $speed_{m}$ and $density_{i}$ in a batch of videos and the music in the Lakh Pianoroll Dataset.
We separate the value range of $speed_{m}$ to 16 classes, the same as the number of classes of $density$, based on the corresponding percentage of density levels.
For example, when there are $5\%$ bars with a $density=16$ in the training set, we classify the top $5\%$ $speed$ as $density=16$. 
Since the $m$th video clip has the same length as the $i$th bar, we replace the $density_{i}$ with classified $speed_{m}$ in inference stage to build the relation, as discussed in Sec. \ref{sec:control}.

\subsection{Motion Saliency and Simu-note Strength} \label{sec:strength}

The \textbf{\textit{motion saliency}} at the $t$th frame is calculated as the average positive change of optical flow of all directions between two consecutive frames.

We then obtain a series of \textbf{\textit{visual beats}}~\cite{visbeat} by selecting frames with both local-maximum motion saliency and a near-constant tempo.
Each visual beat is a binary tuple $(t, s)$ where $t$ is its frame index and $s$ is its saliency. 
As shown in Fig.~\ref{fig:firstpage}, $s$ will have a large value when a sudden visible change occurs.

As shown in Fig.~\ref{illustration}, we define the \textbf{\textit{simu-note strength}} as the number of notes in it:
\begin{equation}
    strength_{i,j,k} = \lvert \emph{simu-note}_{i,j,k} \rvert.
\end{equation}


Intuitively, simu-note strength denotes the richness of an extended chord or harmony, giving the audience a rhythmic feeling along with its progression. 
The higher simu-note strength it has, the more auditory impact the audience will feel. 
We establish a positive correlation between visual beat saliency and simu-note strength so that a distinct visual motion will be expressed by a clear music beat, making the video more rhythmic.

\section{Controllable Music Transformer}\label{sec:cmt}

On top of the above established video-music rhythmic relations, we propose a transformer-based approach to generate background music for given videos, termed Controllable Music Transformer (CMT).
The overall framework is shown in Fig. \ref{framework_fig}. 
We extract rhythmic features from both video and MIDI, which is indicated in the above section.
In the training stage, only rhythmic features from MIDI are included.
In the inference stage, we replace the rhythmic feature with that from the video to perform controllable music generation.

\subsection{Music Representation} 
\label{sec:representation}

\begin{figure}[t]
  \centering
  \includegraphics[width = 0.95\linewidth]{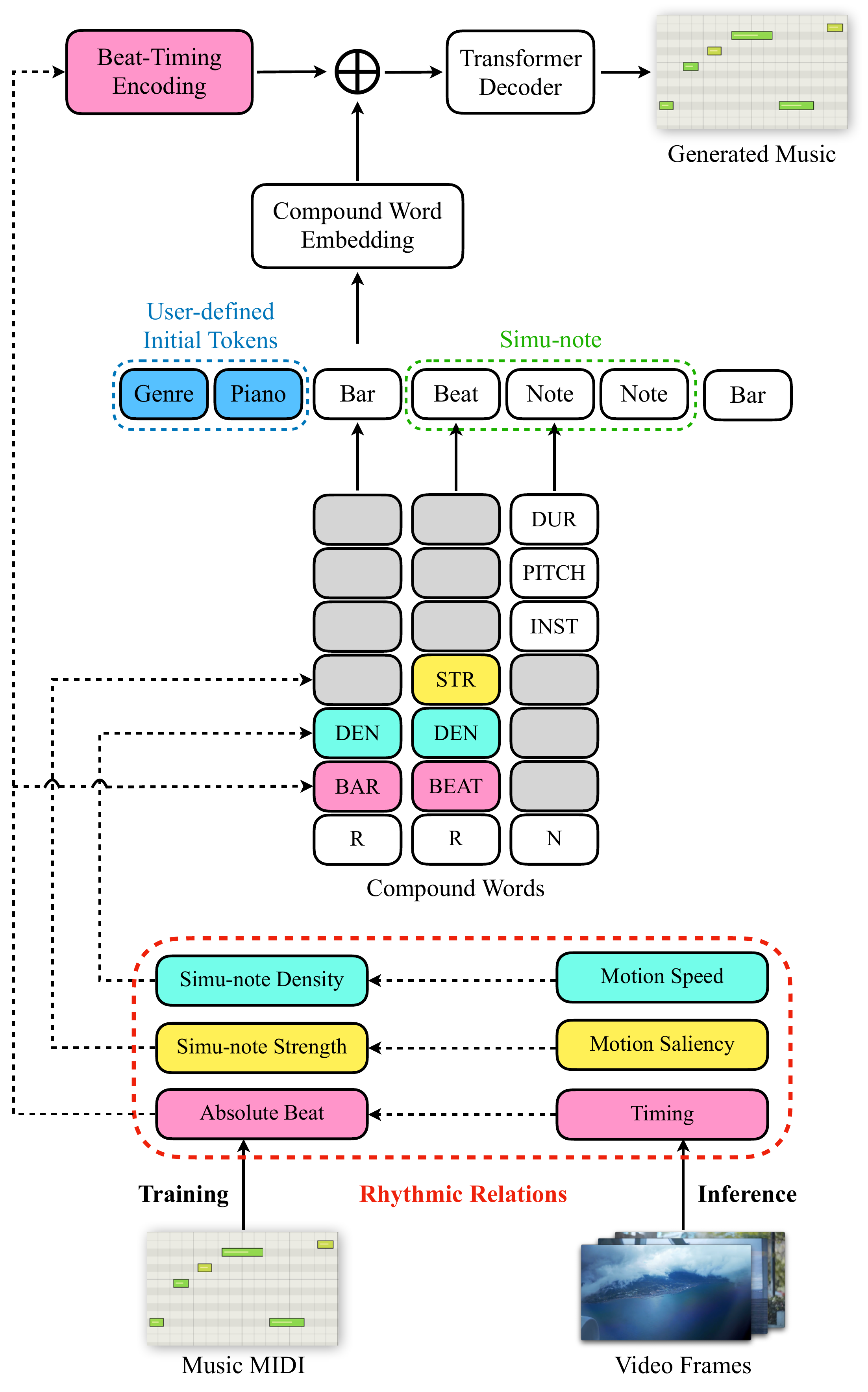} 
  \caption{Illustration of the proposed CMT framework. We extract rhythmic features from MIDI music (in training) or the video (in inference), and construct compound words as the representation of music tokens. The compound word embedding is then combined with beat-timing encoding, and fed into the transformer for sequence modeling.}
  \label{framework_fig}
\end{figure}

We design a music token representation for controllable multi-track symbolic music generation. Inspired by PopMAG~\cite{popmag} and CWT~\cite{cp}, we group related attributes into a single token to shorten the sequence length. 

As shown in Fig.~\ref{framework_fig}, we consider seven kinds of attributes: \emph{type, beat/bar, density, strength, instrument, pitch} and \emph{duration}.
We separate those attributes into two groups: 
a rhythm-related group (denoted as R in Fig. \ref{framework_fig}) including \emph{beat, density} and \emph{strength}, and a note-related group (denoted as N in Fig.~\ref{framework_fig}) including \emph{pitch, duration} and \emph{instrument type} that the note belongs to.
Then we use the \emph{type} attribute (the R/N row of Fig. \ref{framework_fig}) to distinguish those two groups.
To make computational modeling feasible, we set note-related attributes to $None$ in the rhythm-related token and vice versa, as shown in Fig. \ref{framework_fig} with the blank attribute. 
Each rhythm token has a \emph{strength} attribute to indicate the number of following note tokens. 
Besides, the \emph{density} attribute is monotonically decreasing in each bar, indicating the number of remaining simu-note in that bar.
The embedding vector for each token is calculated as follows:
\begin{equation}
    p_{i, k} = \text{Embedding}_{k} \left ( w_{i, k} \right ), k=1,...,K,
    \label{equ:CWT1}
\end{equation}
\begin{equation}
    x_{i} = W_{in}\left[p_{i, 1} \oplus ... \oplus p_{i, K} \right],
    \label{equ:CWT2}.
\end{equation}
Here $w_{i, k}$ is the input words for attribute $k$ at the $i^{th}$ time step before embedding, and $K$ is the number of  attributes in each token. Here $K=7$. $W_{in}$ is a linear layer to project the concatenated embedded token to $R^{d}$. Here $d=512$ is a pre-defined hyper-parameter for the embedded size. $x_{i} \in R^{d}$ is the embedded token.

Besides, we take the genre and instrument type of each music as initial tokens and apply an independent embedding layer for it. The embedded size is the same as ordinary tokens.

\subsection{Control} \label{sec:control}

\par After finishing training, it is expected that the CMT model has already understood the meaning behind strength and density attributes. 
Thus, we only need to replace those two attributes when appropriate in the inference stage to make the music more harmonious with the given video. 

\textbf{Density replacement.} To make the density of the generated music match the density of motion speed of the video, we replace the density attribute on each bar token with the corresponding video density extracted from the optical flow.
Since the CMT model has already learned the meaning of density in the bar token, it will automatically generate the corresponding number of beat tokens in this bar, making the density controllable.

\textbf{Strength replacement.} Likewise, we take advantage of the information from the visual beats of the video to control the strength of the generated simu-notes. 
If the CMT model predicts a beat token after or at the given visual beat, we replace that beat token with the given visual beat and its strength. Then, the CMT model will automatically predict the corresponding number of note tokens in this beat, making strength controllable.

\textbf{Hyper-parameter $C$ for control degree.}
We also need to consider the trade-off between the compatibility of music with the video and its melodiousness. 
This is because the more constraint we put in inference, the more unnatural music we will get. 
To deal with this problem, we design a hyper-parameter $C$ to indicate the control degree on the generated music. 
The larger the $C$ is, the more constraint will be added in inference.
This means we will get a piece of music from scratch when $C$ equals 0 and get full compatibility when $C$ equals 1.
Users can set $C$ according to their own needs.


\textbf{Beat-timing encoding.} In order to leverage the time or length information from the video, we add a beat-timing encoding on the token embedding both in training and inference.
This design tells the CMT model when to start a prologue and when to predict an EOS token.
Beat-timing encoding indicates the ratio of the current beat number to the total beat number in the given video. 
Specifically, we divide the value range of that ratio into $M$ bins (classes) with the same width and use a learnable embedding layer to project it to the same dimension as the token embedding. 
Then we add them together to form the input for our CMT model.

Assuming $i$ is the index of the token, $beat_i$ is the beat number generated at the current step and $N_{beat} = f_{beat}(T)$ is the total beat number of the video, and $T$ is the total frame number of the video, we compute beat-timing encoding according to the equations below:
\begin{equation}
    t_{i} = \text{Embedding}_{t} \left (round \left (M\frac{beat_i}{N_{beat}}\right )\right),
    \label{equ:time_encoding1}
\end{equation}
\begin{equation}
    \vec{x_{i}} =  x_{i} + BPE + t_{i},
    \label{equ:time_encoding3}
\end{equation}
where $t_i \in R^{d}$ is beat-timing encoding for the token $i$, $x_{i}$ is the embedding vector explained in Equation (\ref{equ:CWT2}), and $BPE$ is beat-based position encoding explained in Equation (\ref{BPE_1}) and (\ref{BPE_2}). We set $M=100$ to separate each video into 100 bins. $\vec{x_{i}}$ is the final input fed into the CMT model.

Moreover, instead of using the traditional position encoding, we introduce a beat-based position encoding for each token.
In particular, each token within the same beat will get the same position encoding. 
It is in line with the semantic information of the music sequence since multiple notes in the same beat will be converted to the same audio segment eventually, regardless of their order in the sequence.

Beat-based position encoding of the $i$-th beat $BPE$ is computed as follows:
\begin{equation}
 BPE(beat_{i}, 2n)=sin(\frac{beat_{i}}{10000^{2n/d_{model}}})
 \label{BPE_1}
\end{equation}
\begin{equation}
 BPE(beat_{i}, 2n+1)=cos(\frac{beat_{i}}{10000^{2n/d_{model}}})
 \label{BPE_2}
\end{equation}
where $d_{model} = 512$ is the model hidden size, and  $ n \in \left [0,...,d_{model}/2 \right)$ is the index of $d_{model}$. Beat-based position encoding will be added on each embedding vector $x_{i}$ in Equation (\ref{equ:time_encoding3}) eventually.

\textbf{Genre and instrument type.} 
In our method, there are 6 genres (Country, Dance, Electronic, Metal, Pop, and Rock) and 5 instruments (Drums, Piano, Guitar, Bass, and Strings). 
We take each from them respectively as the initial token for our CMT model. 
Users can choose different genres and instruments by using different initial tokens in the inference stage to generate the background music that matches up with the emotion of the video.

The above-mentioned controlling strategies will be combined together in the inference stage. The more detailed inference algorithm is illustrated in Algorithm \ref{alg:A}.

\begin{algorithm}
\caption{Inference stage}
\label{alg:A}
\begin{algorithmic}
\STATE {Set initial genre and instrument tokens} 
\REPEAT 
\STATE Predict \emph{next token} with given beat-timing from the video based on sampling strategy
\IF {\emph{next token} is bar}
\STATE Replace its density attribute with prob of C
\ENDIF
\IF {\emph{next token} is after visual beat}
\STATE Replace \emph{next token} with visual beat and its strength with prob of C
\ENDIF
\STATE {Append \emph{next token} to \emph{music token list}}
\UNTIL{EOS token predicted}
\RETURN {\emph{music token list}}
\end{algorithmic}
\end{algorithm}

\begin{table*}[!htb]
\centering
\begin{tabular}{c|c|c||c|c|c|c|c||c|c|c}
\toprule
\multicolumn{2}{c|}{Model} & Data & \multicolumn{5}{c||}{Without Control} & \multicolumn{3}{c}{With Control} \\\cline{2-11}
\multicolumn{2}{c|}{No.} & - & 1 & 2 & 3 & 4 & 5 & 6 & 7 & 8 \\\cline{2-11}
\hline
\multicolumn{2}{c|}{Density} & - & - & $\circ$ & - & - & $\circ$ & $\bullet$ & $\circ$ & $\bullet$ \\ 
\multicolumn{2}{c|}{Strength} & - & - & - & $\circ$ & - & $\circ$ & $\circ$ & $\bullet$ & $\bullet$  \\ 
\multicolumn{2}{c|}{Beat-timing encoding} & - & - & - & - & $\surd$ & $\surd$  & $\surd$ & $\surd$ & $\surd$  \\
\hline
\multicolumn{2}{c|}{Pitch Histogram Entropy} & 4.452 & 3.634 & 2.998 & 3.667 & 3.573 & 3.617 & 3.496 & 4.044 & \textbf{4.113} \\
\hline
\multicolumn{2}{c|}{Grooving Pattern Similarity} & 0.968 & 0.677 & 0.714 & 0.647 & 0.778 & \textbf{0.810} & 0.773 & 0.678 & 0.599 \\
\hline
\multicolumn{2}{c|}{Structureness Indicator} & 0.488 & 0.219 & 0.227 & 0.215 & 0.223 & 0.241 & \textbf{0.268} & 0.211 & 0.200 \\
\hline
\multicolumn{2}{c|}{Overall Rank~$\downarrow$} & - & 5.000 & 5.000 & 5.333 & 4.000 & \textbf{2.667} & 3.667 & 4.667 & 5.667 \\
\bottomrule
\end{tabular}
\vspace{0.25cm}
\caption{Objective evaluation for each controlling attribute on melodiousness.  - means that the attribute is not added during training. $\circ$ denotes that the attribute is only added in training but not controlled in the inference time. $\bullet$ means that we not only add it during training but also control that attribute with the corresponding rhythmic feature from a given video in inference. $\surd$ denotes that we add beat-timing encoding in both the training and inference stage. See Sec. \ref{sec:objective} for details.}
\label{Objective}
\end{table*}

\subsection{Sequence Modeling}

The sequence of music tokens (as explained in Sec.~\ref{sec:representation}) is fed into the transformer \cite{transformer} model to model the dependency among elements.
We employ the linear transformer \cite{linear_transformer} as our backbone architecture, considering its lightweight and linear-complexity in attention calculation.

The multi-head output module, following the design of \cite{cp}, predicts 7 attributes of each music token in a two-stage way. In the first stage, the model predicts the \emph{type} token by applying a linear projection on the output of the transformer. In the second stage, it uses \emph{type} to pass through 6 feed-forward heads to predict the remaining 6 attributes at the same time. During inference, we adopt stochastic temperature-controlled sampling \cite{curious} to increase the diversity of the generated tokens.

\section{Experiments}

We perform an ablation study on three control attributes on music generation we propose in this work. 
Both objective and subjective evaluations are conducted.
Objective evaluations focus on the quality of the generated music itself, where we generate ten music pieces for each genre using all the instruments in the initial tokens for each video.
Then we calculate the average on each objective metric.
For subjective evaluation, we designed a questionnaire and invited users to evaluate the quality of the generated music as well as its compatibility with the corresponding video. 

\subsection{Dataset}

We adopt the Lakh Pianoroll Dataset (LPD)~\cite{MuseGan} to train our CMT model. LPD is a collection of 174,154 multi-track pianorolls derived from the Lakh MIDI Dataset (LMD)~\cite{LMD}. We use the \textit{lpd-5-cleansed} version of LPD, which goes through a cleaning process and has all tracks in a single MIDI file merged into five common categories (Drums, Piano, Guitar, Bass, and Strings). 
We then select 3,038 MIDI music pieces from \textit{lpd-5-cleansed} as our training set. The selected pieces fall into six genres (Country, Dance, Electronic, Metal, Pop, Rock) from the tagtraum genre annotations~\cite{schreiber2015improving}.

\subsection{Implementation Details}
Following the design in \cite{cp}, we choose the embedding size for each attribute based on its vocabulary size, \emph{i.e}. $(32, 64, 64, 64, 512, 128, 32)$ for (\emph{type, beat, density, strength, pitch, duration, and instrument}) attributes respectively. Those embedded attributes are concatenated together and projected to model hidden size in Equation (\ref{equ:CWT2}). 

As for the model settings, we use 12 self-attention layers, each with 8 attention heads. The model hidden size and inner layer size of the feed-forward part are set to 512
and 2,048, respectively. The dropout rate in each layer is set to 0.1. The input sequence length is padded to 10,000 with $\langle EOS \rangle$ token.

We set the initial learning rate to 1e-4 and use Adam as the optimizer. We train our model for 100 epochs on the LPD dataset, taking approximately 28 hours on 4 RTX 1080Ti GPUs. The objective metrics are computed with MusDr \cite{jazz_transformer}.

\subsection{Objective Evaluation} \label{sec:objective}

Here we analyze the contribution of each of the controllable attributes. We adopt some of the objective metrics from \cite{jazz_transformer}, including \textbf{Pitch Class Histogram Entropy} that assesses the music's quality in tonality, \textbf{Grooving Pattern Similarity} that measures the music's rhythmicity, and \textbf{Structureness Indicator} that measures the music's repetitive structure. 
Note that the overall quality is not indicated by how high or low these metrics are but instead by their closeness to the real data. 
Finally, we sort each metric among the eight models in Tab. \ref{Objective} and take the mean of the ranking results among the three metrics mentioned above as the final criterion, namely \textbf{Overall rank}.

For each model with control in Tab. \ref{Objective}, we generate 10 MIDI for each genre using all five instruments for each video. To make a fair comparison, we generate the same number of MIDI with the same length of the controlled counterparts for each model without control.

In Tab. \ref{Objective}, we first statistically analyze the objective metrics on the LPD dataset. 
The generated music should be close to music in the dataset to be more natural regarding the adopted metrics. 
Then we train a model without our proposed three rhythmic features, \emph{i.e.,} experiment No.1 in Tab. \ref{Objective}, serving as a baseline model for comparison.
Then we add each rhythmic feature one by one, \emph{i.e.,} experiment No. 2, 3, 4 in  Tab. \ref{Objective}. 
Density and beat-timing encoding help improve the overall structures of the music. 
Strength makes it easy for the model to learn the combination of different pitch classes to form a simu-note, leading to a better pitch histogram entropy.
When we combine those three proposed rhythmic features, as shown in experiment No.5, we take all the advantages and get a higher score on each metric than the baseline model, indicating an improvement in overall melodiousness. However, when we try to control those attributes with the given video, \emph{i.e.,} experiment No. 6, 7, 8 as shown in Tab. \ref{Objective}, we observe a degeneration on structures of the music. 
It is reasonable since we force the generated music to align with the rhythm from the video. Thus, in Sec. \ref{sec:subjective}, we conduct a user study on hyper-parameter $C$, in order to find the trade-off of the structures degeneration and the music-video compatibility.

To sum up, the highest overall rank for experiment No. 5 demonstrates that the rhythmic features we extract from the music not only make the controlling workable in inference but also improve the overall melodiousness of the generated music because the extracted density and strength prompt the CMT model to learn the underlying pattern of the rhythm of the background music.
In other words, the rhythmic feature makes it easy for the network to converge and thus improves the structures of the music.

\subsection{Subjective Evaluation}\label{sec:subjective}

The best way to evaluate a music generation model today remains using user study. 
Besides, objective metrics do not consider the extent to which the video and music are matched.
Therefore, we design a questionnaire for subjective evaluation of our model and invite 36 people to participate. 13 among them have related experience in music or basic understanding of music performing and are considered professionals. Each participant is asked to listen to several pieces of music (in random order) corresponding to one input video, rate based on subjective metrics introduced below, and rank them based on their preferences. As the music can be long, the questionnaire may take around 10 minutes to complete.

We select several subjective metrics \cite{cp} to evaluate the melodiousness of our music: 1) \textbf{Richness}: music diversity and interestingness; 2) \textbf{Correctness}: perceived absence of notes or other playing mistakes. (\emph{i.e.}, weird chords, sudden silence, or awkward usage of instruments); 3) \textbf{Structuredness}: whether there are structural patterns such as repeating themes or development of musical ideas.

Moreover, in terms of the compatibility of the music with the given video, we choose the following metrics for evaluation: 
1) \textbf{Rhythmicity}: how much the rhythm of the generated music matches with the motion of the video. For example, an intense sports vlog with large movements should be matched up with fast-paced music.
A clam and smooth travel vlog with gentle movements should be matched up with slow-paced music. 
2) \textbf{Correspondence}: how much the major stress or the boundary of the generated music matches with the video boundary or the visual beat. For instance, rhythmic motion, such as dancing, and some obvious video boundaries should be accompanied by major stress to improve musicality.
3) \textbf{Structure Consistency}: the start and the end of the generated music should match up with those of the video. Similarly, music and video should both have a prologue, epilogue, and episode, so those structures should be matched to make the video more harmonious.

To take all those subjective metrics into account and give a comprehensive evaluation of the generated background music, we ask participants to rank those videos based on the overall quality, and then we take the mean of the rank as the final result, namely \textbf{Overall rank}.

We first experiment on different levels of hyper-parameter $C$ to choose an appropriate value for the trade-off issue between compatibility with the video and melodiousness of the generated music. We choose a given video, and for each $C$ value, we run the inference stage and generate one music clip. All music clips are included in the questionnaire to be evaluated by the participants. The result is shown in Tab. \ref{Subjective_C}. Although when $C = 1.0$, we get better compatibility between the video and the music, it is detrimental to the melodiousness of the music, especially on the correctness metrics, leading to a lower overall rank. It is reasonable since the constraint on the rhythm will force the model to generate some relatively unnatural notes. Considering the overall rank, we eventually take $C = 0.7$ as our pre-defined hyper-parameter.

\begin{table}[t]
\centering
\begin{tabular}{c|c|c|c}
\toprule
Model & Baseline & Matched & Ours   \\ 
\hline
Melodiousness $\uparrow$ & 3.4 & \textbf{4.0} & 3.8 \\
\hline
Compatibility $\uparrow$ & 3.4 & 3.7 & \textbf{3.9} \\
\hline
Overall Rank $\downarrow$ & 2.3 & 1.9 & \textbf{1.8} \\
\bottomrule
\end{tabular}
\vspace{0.25cm}
\caption{Subjective evaluation on melodiousness and compatibility with the video with $C=0.7$. Our music reaches comparable performance compared with matched data from training set, especially in compatibility with the video.}
\label{Subjective}
\end{table}

\begin{figure*}[!t]
    \centering
    \subfigure[barbeat loss]{
        \includegraphics[width=1.6in, height=1.1in]{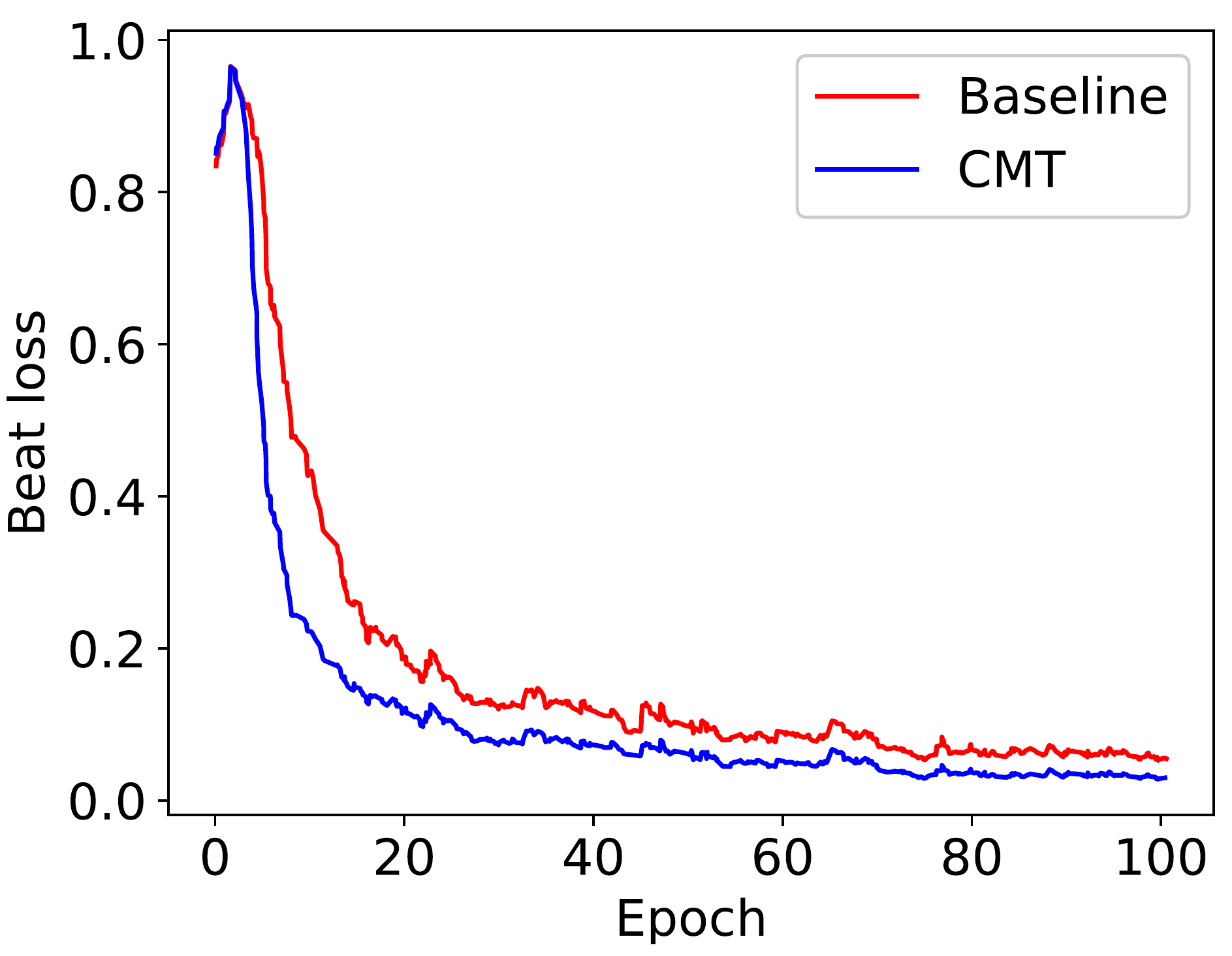}
        \label{barbeat_loss}
    }
    \hspace{-0.0cm}
    \subfigure[duration loss]{
	    \includegraphics[width=1.6in, height=1.1in]{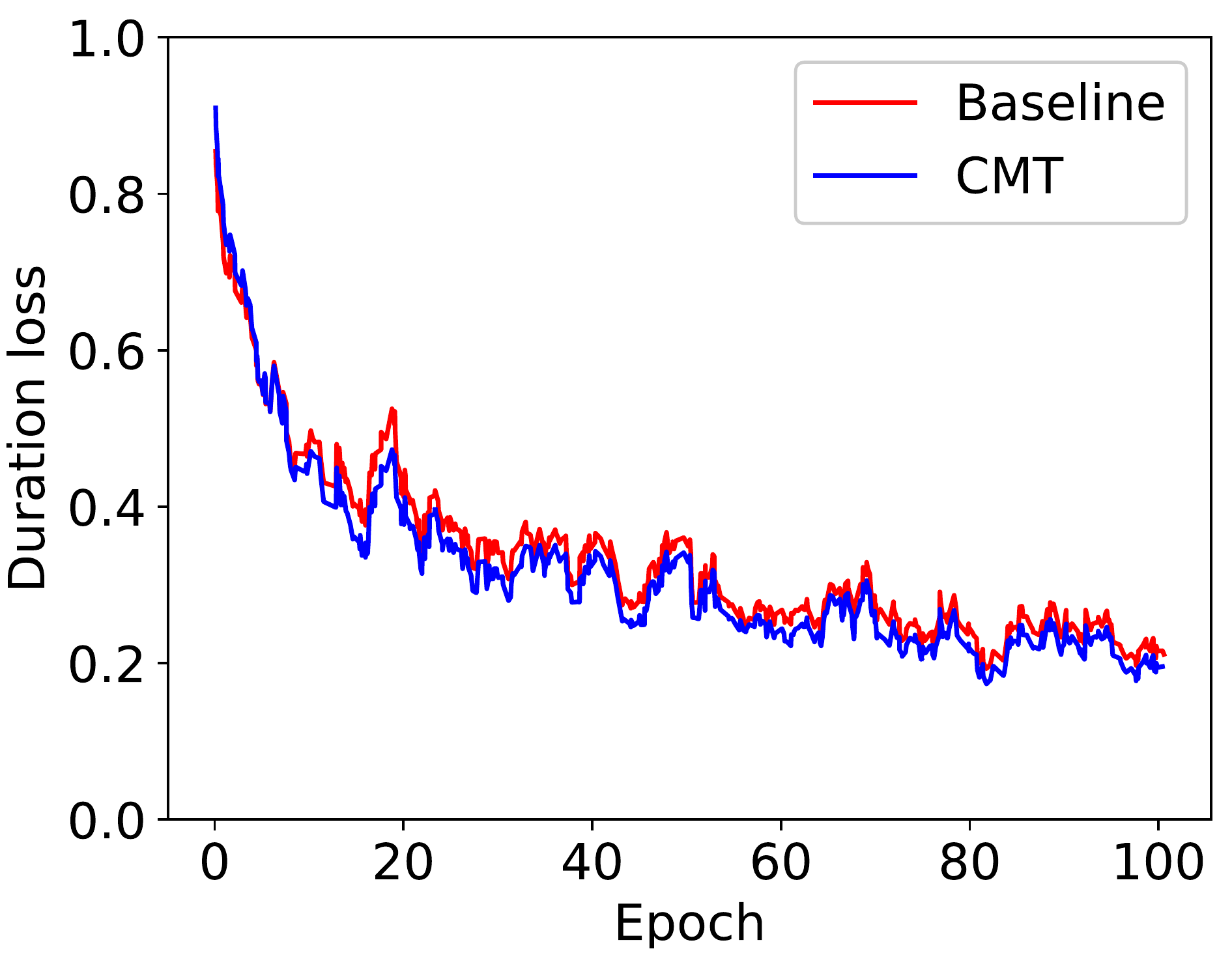}
	    \label{duration_loss}
    }
    \hspace{-0.0cm}
    \subfigure[pitch loss]{
        \includegraphics[width=1.6in, height=1.1in]{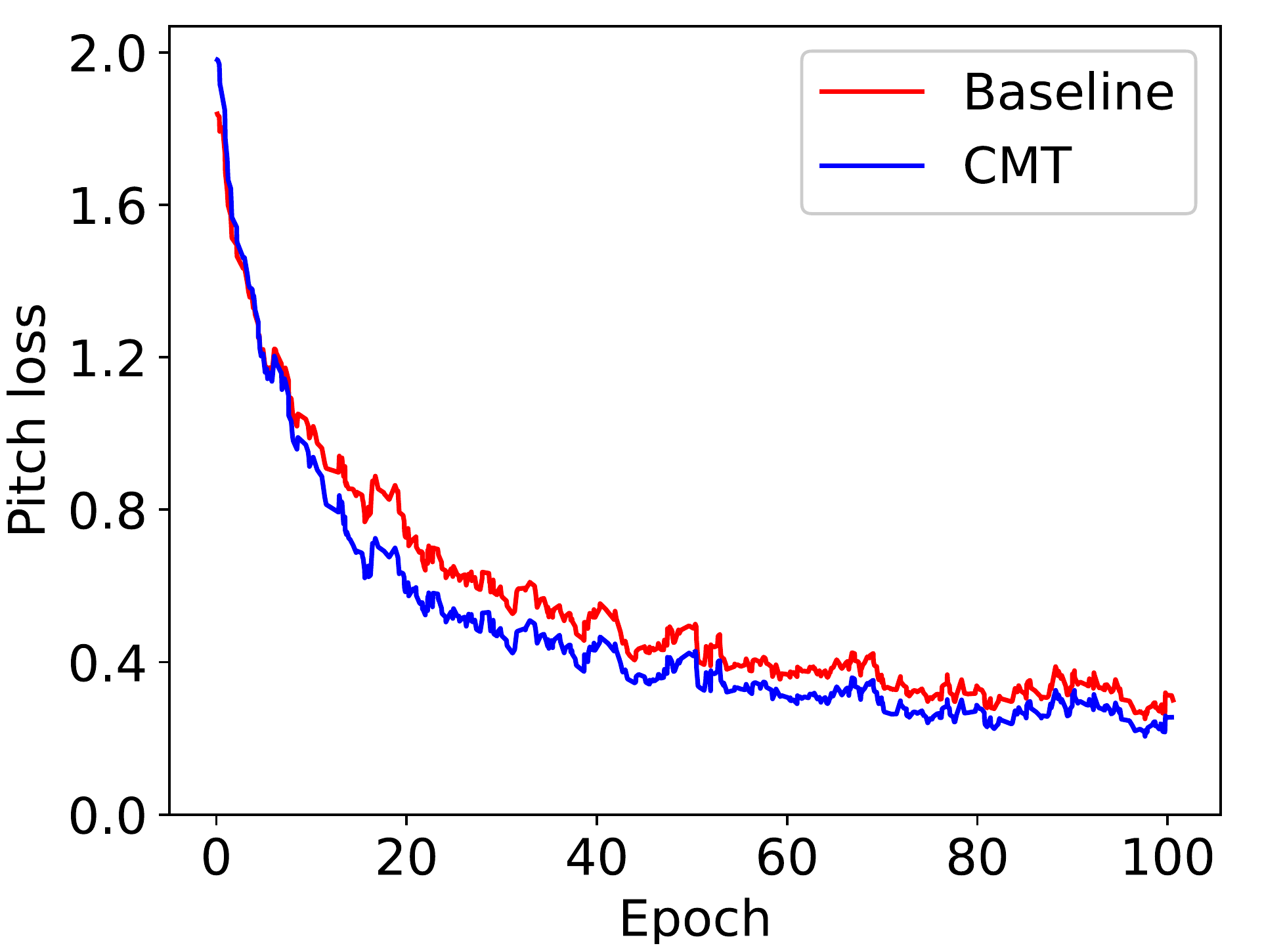}
        \label{pitch_loss}
    }
    \hspace{-0.0cm}
    \subfigure[instrument loss]{
        \includegraphics[width=1.6in, height=1.1in]{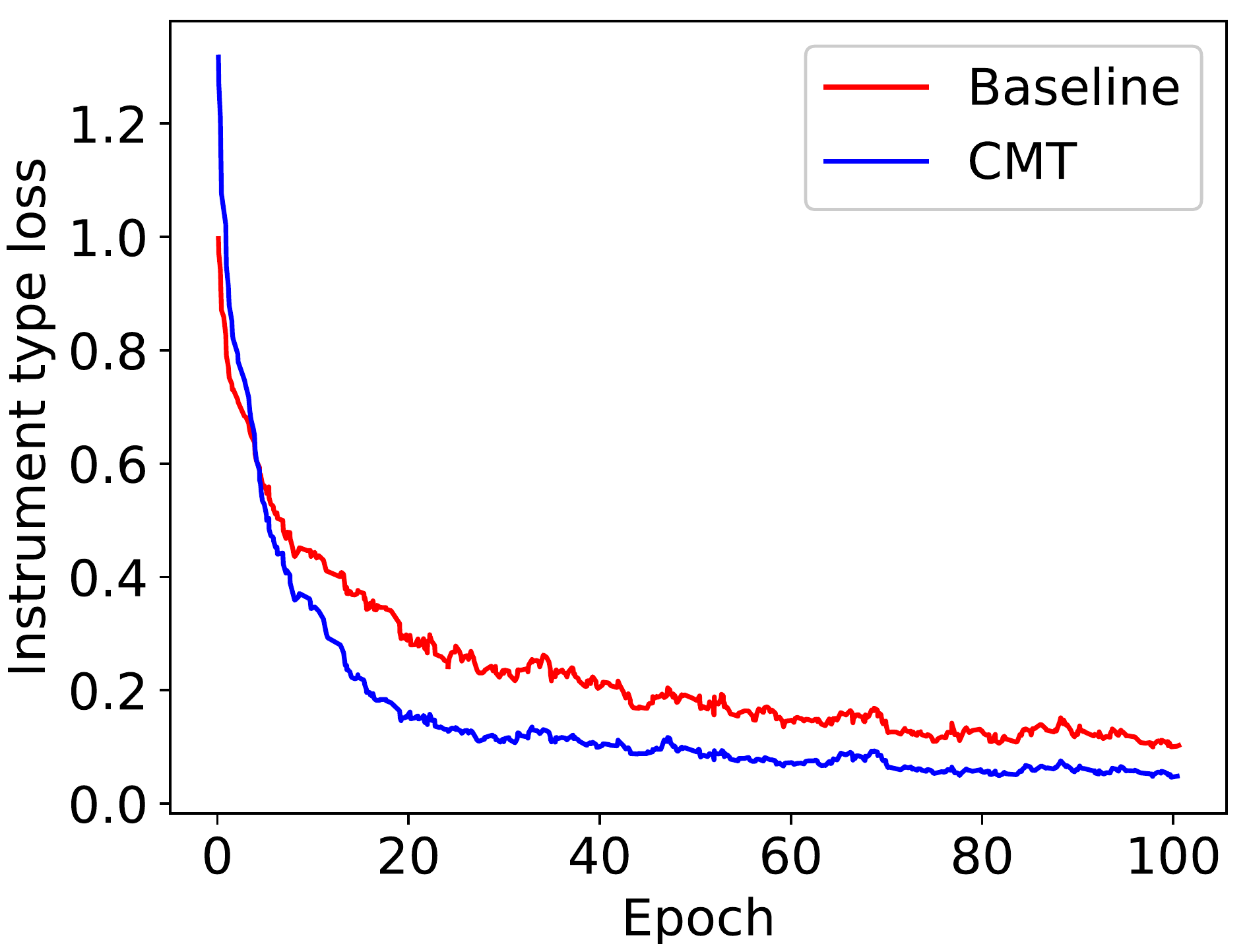}
        \label{instrument_loss}
    }
    \vspace{-0.4cm}
    \caption{The loss curve for the baseline model and the CMT model. The blue line is the baseline model, and the red line is our proposed CMT model. Our method shows an increased converging speed with the help of density, strength, and beat-timing encoding. For a better demonstration, we perform exponential smoothing on each loss with $\alpha = 0.9$.}
    \label{loss_figure}
    \vspace{0.2cm}
\end{figure*}

Then, we evaluate our music in terms of melodiousness and compatibility with the given video. The baseline model is the one without using any controllable attributes. Moreover, we design an algorithm to match a video with music from the training set based on our proposed rhythmic relations. Specifically, given a video and a music piece, we calculate their matching score ($MS$) as:
\begin{equation}
    MS(\mathbf{d^m}, \mathbf{d^v}, \mathbf{s^m}, \mathbf{s^v}) = \frac{1}{MSE(\mathbf{d^m}, \mathbf{d^v}) + MSE(\mathbf{s^m} \odot \mathbb{1}(\mathbf{s^v}), \mathbf{s^v})}.
\end{equation}
Here, $\mathbf{d^m}$ and $\mathbf{d^v}$ are simu-note density extracted from the music and the video and are truncated to the same size. Likewise, $\mathbf{s^m}$ and $\mathbf{s^v}$ are extracted simu-note strength with truncation. $MSE$ is the mean squared error, $\odot$ denotes the Hadamard product, $\mathbb{1}(\cdot)$ maps each positive element to 1 and non-positive element to 0. We then manually select one from the top-5 matched music based on the video style.

We choose three video clips from different categories (edited, unedited, and animation video) and provide the generated music of \emph{ours}, \emph{baseline} and \emph{matched} in the questionnaire. They are randomly shuffled, so the participants do not know which one is generated by the model and which one is selected from the dataset.

Tab. \ref{Subjective} demonstrates that our generated background music even surpasses the matched music overall. The matched music shows better compatibility than our baseline, indicating that our proposed rhythmic relation is valuable and beneficial for the overall musicality of the video. Although the melodiousness of our composed music is still below the real one in the training set, the excellent compatibility compensates for those weaknesses, making the generated background music more suitable than human-made music.

\subsection{Controlling Accuracy}

\begin{table}[t]
\centering
\begin{tabular}{c|c|c|c|c|c}
\toprule
\multicolumn{2}{c|}{\multirow{2}*{Metrics}} & \multicolumn{4}{c}{\textbf{$C$}}  \\ 
\cline{3-6}
\multicolumn{2}{c|}{~} & 0.0 & 0.3 & 0.7 & 1.0  \\ 
\hline
\multirow{3}*{Melodiousness} & Richness~$\uparrow$  & 3.6 & 3.4 & \textbf{3.8} & 3.7 \\
\cline{2-6}
~ & Correctness~$\uparrow$  & 3.2 & \textbf{3.7} & \textbf{3.7} & 2.8 \\
\cline{2-6}
~ & Structuredness~$\uparrow$ & \textbf{3.6} & \textbf{3.6} & \textbf{3.6} & 3.3 \\
\hline
\multirow{3}*{Compatibility} & Rhythmicity~$\uparrow$ & 3.2 & 3.5 & \textbf{3.7} & \textbf{3.7} \\
\cline{2-6}
~ & Correspondence~$\uparrow$ & 2.6 & 3.3 & 3.7 & \textbf{4.1}  \\
\cline{2-6}
~ & Structure Consistency~$\uparrow$ & 2.9 & \textbf{3.9} & 3.8 & 3.8 \\
\hline
\multicolumn{2}{c|}{Overall rank~$\downarrow$}  & 3.1 & 2.2 & \textbf{2.1} & 2.6  \\
\bottomrule
\end{tabular}
\vspace{0.25cm}
\caption{Subjective ablation study for different $C$ values on melodiousness and compatibility with the video. We observe that higher $C$ leads to higher compatibility, while lower $C$  leads to better melodiousness. The overall performance reaches its peak when $C$ is set to 0.7.}
\label{Subjective_C}
\vspace{-0.25cm}
\end{table}

\begin{table}[t]
\centering
\begin{tabular}{c|ccc}
\toprule
Attribute & Density & Strength & Time \\
\midrule
Control Error & 0.107 & 0.001 & 0.028 \\
\bottomrule
\end{tabular}
\vspace{0.25cm}
\caption{The error rate for density, strength, and time control. Our method demonstrates impressive performance in controlling the three rhythmic features of the music.}
\label{controllable_table}
\end{table}

We evaluate the accuracy of the proposed three controllable attributes. We recalculate those three attributes in the music and take the L2 distance between the rhythmic feature from the video and our music as the control error. Then, errors of density, strength and time are normalized by the average number of simu-notes per bar, average number of notes per simu-note and the total video time, respectively, to eventually form the error rates. As shown in Tab. \ref{controllable_table}, our results are impressive. The control error for music density is around 0.1 while the average number of simu-notes per bar is 9.9, which means in each bar the beat number will approximately fluctuate only one beat with the given video density. The strength control error shows that the majority of the simu-notes will have the exact same number of notes as the given video strength.

\subsection{Visualization}

We visualize the loss curves for the rhythm-related attributes and the note-related attributes on both the baseline model and our CMT model. The results are shown in Fig. \ref{loss_figure}. It is obvious that our CMT model has a faster converging process on each attribute, especially on the \emph{beat} attribute. That is to say, our extracted rhythmic feature make it easy for the model to grasp the crucial knowledge in music, resulting in a more fetching generated music.

\section{Conclusion}
 In this paper, we address the unexplored task -- \emph{video background music generation}. We first establish three rhythmic relations between video and background music. We then propose a Controllable Music Transformer (CMT) to achieve local and global control of the music generation process. 
 Our proposed method does not require paired video and music data for training while generates melodious and compatible music with the given video.
 Future studies may include exploring more abstract connections between visual and music (e.g., emotion and style), utilizing music in the waveform,  and adopting unsupervised audiovisual representation learning from paired data.
\begin{acks}
This research is supported in part by National Natural Science Foundation of China (Grant 61876177), Beijing Natural Science Foundation (4202034), Fundamental Research Funds for the Central Universities.
\end{acks}

\bibliographystyle{ACM-Reference-Format}
\balance
\bibliography{sample-base}










\end{document}